\documentclass[12pt,preprint]{aastex}

\begin{document}

\title{Photometric Properties of the Near-contact Binary GW Geminorum}
\author{Jae Woo Lee$^{1}$, Seung-Lee Kim$^{1}$, Chung-Uk Lee$^{1}$, Ho-Il Kim$^1$, Jang-Ho Park$^{1}$, So-Ra Park$^{1}$, and Robert H. Koch$^{2}$}
\affil{$^1$Korea Astronomy and Space Science Institute, Daejeon 305-348, Korea}
\email{jwlee@kasi.re.kr, slkim@kasi.re.kr, leecu@kasi.re.kr, hikim@kasi.re.kr, pooh107162@kasi.re.kr, bluemoon08@kasi.re.kr}
\affil{$^2$Department of Physics and Astronomy, University of Pennsylvania, Philadelphia, USA}
\email{rhkoch@earthlink.net}

\begin{abstract}
New multiband CCD photometry is presented for the eclipsing binary GW Gem; the $RI$ light curves are the first
ever compiled. Four new minimum timings have been determined. Our analysis of eclipse timings observed during
the past 79 years indicates a continuous period increase at a fractional rate of $+$(1.2$\pm$0.1)$\times$10$^{-10}$,
in excellent agreement with the value $+$1.1$\times$10$^{-10}$ calculated from the Wilson-Devinney binary code.
The new light curves display an inverse O'Connell effect increasing toward longer wavelengths.
Hot and cool spot models are developed to describe these variations but we prefer a cool spot
on the secondary star. Our light-curve synthesis reveals that GW Gem is in a semi-detached, but near-contact,
configuration. It appears to consist of a near-main-sequence primary star with a spectral type of about A7 and
an evolved early K-type secondary star that completely fills its inner Roche lobe. Mass transfer from the secondary
to the primary component is responsible for the observed secular period change.
\end{abstract}

\keywords{Stars}

\section{INTRODUCTION}

GW Gem ($\rm BD+27^{o}1494$, $V$=$+$10.485, $B-V$=$+$0.27) was announced by Hoffmeister (1949) to be an Algol-type variable.
Subsequently, eclipse timings of the system have been reported assiduously by numerous workers, and history is now long
enough to understand the binary's period behavior. Light curves have been made only by Broglia \& Conconi
(1981a, hereafter B\&C) from photoelectric observations during the seasons of 1978 and 1979.  Unfortunately,
the comparison star ($\rm BD+28^{o}1494$) used in their observations turned out to be a $\delta$ Scuti-type variable
with a peak-to-peak amplitude of about 0.035 mag (Broglia \& Conconi 1981b). B\&C analyzed their $BV$ light curves,
and concluded that GW Gem is a semi-detached system with the secondary star filling its inner Roche lobe. They also
suggested that the period of the binary system be considered constant. Most recently, Shaw (1994) included the system
in his second catalog of near-contact binaries but with an unknown subclass. In this article, we report and analyze
multiband light curves and present the first detailed analysis of the $O$--$C$ diagram.

\section{OBSERVATIONS}

Our CCD photometry of GW Gem was performed on 13 nights from 20 December 2007 through 1 March 2008 in order to obtain
multicolor light curves. The observations were taken with a SITe 2K CCD camera and a $BVRI$ filter set attached
to the 61-cm reflector at Sobaeksan Optical Astronomy Observatory (SOAO) in Korea. The instrument and reduction method
are the same as those described by Lee et al. (2007). Since the field of view of an individual CCD image was large enough
to observe a few tens of nearby stars simultaneously, we monitored many of them on each frame. 2MASS 07530545+2716551
($V$=$+$11.573, $B-V$=$+$0.534 from the NOMAD Catalogues; Zacharias et al. 2005) was used as a comparison star and
no brightness variation of it was detected against measurements of the other monitoring stars during our observing runs.
A total of 2,078 individual observations was obtained among the four bandpasses (508 in $B$, 511 in $V$, 532 in $R$,
and 527 in $I$) and a sample of them is listed in Table 1. The light curves are plotted in Figure 1 as differential magnitudes
of ${\rm m_{var}}-{\rm m_{comp}}$ {\it versus} orbital phase.

\section{PERIOD STUDY}

From our CCD observations, times of minimum light in each filter have been determined with the method of
Kwee \& van Woerden (1956). Weighted mean timings from these determinations and their errors are given in Table 2, together
with all other photoelectric and CCD timings. The second column gives the standard deviation of each timing. In all,
there were assembled 155 eclipse timings (48 photographic plate, 80 visual, 2 photographic, 10 photoelectric and 15 CCD)
from the literature and from our CCD measures. Except for the photoelectric and CCD minima, the other timings were
extracted from the modern database published by Kreiner et al. (2001).  An error for each method was assigned: $\pm$0.0230 d
for photographic plate, $\pm$0.0059 d for visual, $\pm$0.0022 d for photographic, and $\pm$0.0012 d for photoelectric and
CCD minima. Relative weights were then scaled from the inverse squares of these values (Lee et al. 2007).

First of all, we constructed an $O$--$C$ diagram for GW Gem using the light elements of Kreiner et al.:
\begin{equation}
 C_1 = \mbox{HJD}~2,425,645.5748 + 0.65944433E.
\end{equation}
The resulting $O$--$C_{1}$ residuals calculated with equation (1) are listed in the fourth column of Table 2 and drawn
in the upper panel of Figure 2, where the timings are marked by different symbols according to observational method.
The general trend of these $O$--$C$ residuals from 1929 to 2008 is a curvilinear pattern. Thus, by introducing all times
of minimum light into a parabolic least-squares fit, we obtained the following quadratic ephemeris:
\begin{equation}
 C_2 = \mbox{HJD}~2,425,645.5823(43) + 0.65944289(26) E  + 3.83(38) \times 10^{-11} E^2.
\end{equation}
The 1$\sigma$-value for the last decimal place of each ephemeris parameter is given in parentheses. The result is
represented as a continuous curve in the upper panel of Figure 2. The $O$--$C_{2}$ residuals from this equation are
given in the fifth column of Table 2 and are plotted in the lower panel of Figure 2.

As can be seen in Figure 2, the upward-parabola ephemeris provides a good fit to the $O$--$C$ residuals and
signifies a continuous period increase at a rate of $+$(4.2$\pm$0.4)$\times$10$^{-8}$ d yr$^{-1}$, corresponding
to a fractional period change of $+$(1.2$\pm$0.1)$\times$10$^{-10}$. Within its error, this agrees well with
the value of $+$1.1$\times$10$^{-10}$ calculated later in this paper with the Wilson-Devinney binary code
(Wilson \& Devinney 1971, hereafter WD). Because our light-curve synthesis shows that GW Gem is in
a semi-detached configuration with the less massive secondary filling its inner Roche lobe,
such a positive quadratic term can be produced by conservative mass transfer from the secondary
to the primary component. From the masses of both components listed in Table 6, we calculate
a modest mass transfer rate of 3.2$\times$10$^{-8}$ M$_\odot$ yr$^{-1}$.

\section{LIGHT-CURVE SYNTHESIS AND SPOT MODELS}

As shown in Figure 1, the morphology of the light curve of GW Gem resembles that of $\beta$ Lyr (dissimilar eclipse depths
and light variability continuous with phase) and therefore indicates a significant temperature difference between
the two components and significant distortion of the photospheres. In order to derive reasonable representations of
the binary system, we analyzed simultaneously both our and the B\&C light curves by applying the WD synthesis code
to all individual observations. For this purpose, we established unit light level at phase 0.75 and used
an observational weighting scheme identical to that for the eclipsing binary RU UMi (Lee et al. 2008a). Table 3 lists
the light-curve sets for GW Gem analyzed in this paper and the standard deviations ($\sigma$) of a single observation.
The noise of the SOAO light curves is somewhat larger than that of the B\&C ones despite the variability of
their comparison star. This is presumably due to the smaller telescope used at SOAO, and/or because our comparison star is
significantly fainter than GW Gem.

The effective temperature of the brighter, and presumably more massive, star was initialized at 7700 K from
Flower's (1996) table, because ($B-V$)=$+$0.24 at Min II (i.e., phase 0.50) as given by B\&C and because of
the small reddening, $E$($B-V$)=$+$0.03, calculated following Schlegel et al. (1998).  There is
a small systematic error made in the temperature assignment because the eclipses are not complete and a minor lune
of the cool star is visible at secondary minimum.  The temperature of the hot star could be as much as 100 K hotter
than has been assigned but not nearly so hot as if the Simbad spectral type of A4 were chosen.  It is appropriate
to regard the hot star envelope as a radiative atmosphere. The logarithmic bolometric ($X$, $Y$) and
monochromatic ($x$, $y$) limb-darkening coefficients were interpolated from the values of van Hamme (1993) and were
used in concert with the model atmosphere option. The light curves were analyzed in a manner similar to those of
the near-contact binaries AX Dra (Kim et al. 2004) and RU UMi (Lee et al. 2008a). In this section and the next one,
subscripts 1 and 2 refer to the stars eclipsed at primary and secondary minima, respectively.

Although a photometric solution of GW Gem was reported by B\&C from the analysis of their own light curves, there is
no spectroscopic mass ratio ($q$) for the system. We therefore conducted an extensive $q$-search procedure
for various modes of the WD code so as to understand generally the geometrical structure and the photometric parameters
of the system (cf. Lee et al. 2008b). In this procedure, the method of multiple subsets (Wilson \& Biermann 1976) was
used to ensure the stability of the result. Only the photometric solutions for mode 5 (semi-detached systems in which
the secondary components fill their inner critical surfaces) are acceptable for GW Gem.  This configuration is also
consistent with mass transfer from the secondary star to the hotter, more massive primary star inferred from
our period study. To confirm the preliminary result, $q$-searches were repeated for each data set individually.
The weighted sums of the squared residuals ($\Sigma W(O-C)^2$ = $\Sigma$) as a function of $q$ are displayed in Figure 3.
Here, circles, squares, and diamonds represent the search results for B\&C, SOAO, and all data sets, respectively.
The optimal solution is close to $q$=0.46.

The value of $q$ was treated as an adjustable parameter in all subsequent calculations to derive photometric solutions.
The best result is listed in Table 4 and plotted in Figure 4, where for clarity individual observations have been compiled
into 200 mean points using bin widths of 0.005 in phase for each filtered light curve. As seen in the figure,
the computed light curves describe the B\&C data satisfactorily, but do not fit the SOAO data around phase 0.25
as well as can be wished. The discrepancies increase toward longer wavelengths and these can be modelled by a cool spot
on the secondary star which must have a convective atmosphere. However, this representation is not unique for one cannot
exclude a possible hot spot on the primary star due to impact from mass transfer between the components. Thus,
two different model spots were postulated to reanalyze the SOAO light curves. We used the unspotted photometric parameters
as initial values and included $x_1$ and $x_2$ as additional free variables. Final results are given in Table 5 and,
within errors, each of these is in excellent agreement with the geometry of the unspotted solution and fits
the SOAO light curves better than the unspotted model. The light residuals from the spot models are plotted in Figure 5
and it can be seen that there is no systematic trend among them. During the evaluation with spot parameters,
we searched for a possible third light source but none was detected in the light-curve analysis.

It is numerically impossible to discriminate between the two spot models but there is more information to be weighed.
It is noteworthy that the older B\&C light curves show no obvious asymmetry but the evidence of the period study is
that mass transfer was ongoing for at least the last 80 years. We could therefore expect an impact hot spot to have
also existed during the 1978 and 1979 seasons but the light curve asymmetry in the B\&C data did not appear.  The easiest way
to reconcile this limited and discontinuous information is to postulate that an inconspicuous impact hot spot has always
existed due to the feeble mass transfer.  Such a spot would be relatively inconspicuous since the hot star is large
with respect to its Roche lobe and the free-fall height onto its photosphere is not great. Kinetic heating from
the transferring gas is therefore modest and the effect in the light curves below the level of precision of the measures.
Finally, for reasons that cannot be known, the cool star had developed a conspicuous cool spot by the time of
the SOAO observations.  Therefore, the cool spot model would be a more reasonable interpretation, and what has been
modelled is the differential effect between a postulated meager hot spot and the more extensive transient cool spot.
Nothing can be known of photometric activity between the older and newer light curves but it would have to be true
that a trifling hot spot continued to be present in that interim.

\section{DISCUSSION AND CONCLUSIONS}

Based on the historical and new photometry, we have presented the first detailed period analysis and light-curve synthesis
for GW Gem. The absolute stellar parameters for the system can be computed from our complete photometric solutions with
the cool-spot model of Table 5 and from Harmanec's (1988) relation between spectral type and physical parameters
(effective temperature, mass, and radius). The dereddened color ($B-V$)$_0$=$+$0.21 and temperature of the primary component
correspond to a normal near-main-sequence star with a spectral type of about A7.  The astrophysical parameters are
listed in Table 6, where the radius ($R_2$) of the secondary star results from the ratio ($r_{2}/r_{1}$=0.856)
of the mean-volume radii for each component. The bolometric corrections (BCs) were obtained from the scaling between
$\log T$ and BC recalculated by Kang et al. (2007) from Flower's table. An apparent visual magnitude of $V$=+10.49
at maximum light (B\&C), appropriate dereddening, and our computed light ratio at phase 0.75 lead to $V_1$=+10.59 and
$V_2$=+13.05 for the primary and secondary stars, respectively. Using the interstellar reddening of $A_V$=0.10,
the calculated value of $V_1$, and the expected value of $M_{V1}$ for the primary star, we calculated an approximate distance
to the system of about 120 pc.

A comparison of the GW Gem parameters with the mass-radius, mass-luminosity, and Hertzsprung-Russell diagrams
(Hilditch et al. 1988) clearly demonstrates that the primary component lies between the zero-age and
terminal-age main-sequence loci, while the secondary is oversized and overluminous for its mass. In these diagrams,
the locations of the two component stars fall amid those of previously-known near-contact binaries. Thus, the system
is a semi-detached and FO Vir-subtype, near-contact binary consisting of a detached main-sequence primary component
with a spectral type of A7 and an evolved secondary component with a spectral type of approximately K1 which fills
its limiting lobe completely. Such a configuration supports the concept of mass transfer from the secondary to
the primary component as indicated by our period analysis. Our results suggest that GW Gem currently is in a state
of broken contact evolving from a contact configuration as predicted by thermal relaxation oscillation theory
(Lucy 1976, Lucy \& Wilson 1979). High-resolution spectroscopy will determine the astrophysical parameters
and evolutionary status of the system better than is possible with photometry alone.

\acknowledgments{ }
The authors wish to thank Prof. Chun-Hwey Kim for his help using the $O$--$C$ database of eclipsing binaries,
and Ms. In Kyung Baik for collecting the B\&C observations on GW Gem. We also thank the staff of
the Sobaeksan Optical Astronomy Observatory for assistance with our observations. We have used the Simbad data base
maintained at CDS many times and appreciate its availability.

\clearpage
\begin{deluxetable}{cccccccc}
\tablewidth{0pt} \tablecaption{CCD photometric observations of GW Gem$\rm ^a$.}
\tablehead{
\colhead{HJD} & \colhead{$\Delta B$} & \colhead{HJD} & \colhead{$\Delta V$} & \colhead{HJD} & \colhead{$\Delta R$} & \colhead{HJD} & \colhead{$\Delta I$}
}
\startdata
455.20948 & $-$1.321  &  455.20594 & $-$1.001  &  455.20703 & $-$0.756  &  455.21267 & $-$0.510   \\
455.22320 & $-$1.317  &  455.21069 & $-$0.985  &  455.21171 & $-$0.779  &  455.21700 & $-$0.526   \\
455.22764 & $-$1.332  &  455.21502 & $-$0.966  &  455.21604 & $-$0.796  &  455.22628 & $-$0.525   \\
455.23177 & $-$1.342  &  455.21933 & $-$0.994  &  455.22032 & $-$0.759  &  455.23065 & $-$0.521   \\
455.23582 & $-$1.302  &  455.22436 & $-$0.995  &  455.22535 & $-$0.771  &  455.23476 & $-$0.517   \\
455.24820 & $-$1.335  &  455.22876 & $-$0.995  &  455.23386 & $-$0.742  &  455.24262 & $-$0.502   \\
455.26056 & $-$1.320  &  455.23685 & $-$0.967  &  455.24172 & $-$0.748  &  455.24656 & $-$0.510   \\
455.26953 & $-$1.288  &  455.24079 & $-$0.977  &  455.24564 & $-$0.766  &  455.25500 & $-$0.521   \\
455.27716 & $-$1.317  &  455.25708 & $-$0.987  &  455.25016 & $-$0.779  &  455.26343 & $-$0.501   \\
455.28098 & $-$1.301  &  455.26159 & $-$0.993  &  455.25409 & $-$0.754  &  455.26724 & $-$0.507   \\
\enddata
\tablenotetext{a}{2,454,000 is to be added to each HJD entry}
\tablecomments{This sample is shown for guidance regarding form and content. The table is presented
 in its entirety in the electronic edition of Publications of the Astronomical Society of the Pacific.}
\end{deluxetable}

\begin{deluxetable}{llrrrccl}
\tabletypesize{\small}
\tablewidth{0pt}
\tablecaption{Observed photoelectric and CCD timings of minimum light for GW Gem.}
\tablehead{
\colhead{HJD} & \colhead{Error} & \colhead{Epoch} & \colhead{$O$--$C_1$} & \colhead{$O$--$C_2$} & \colhead{Method} & \colhead{Min} & References \\
\colhead{(2,400,000+)} & & & & & & & }
\startdata
43,514.53438   &  $\pm$0.00003  &  27,097.0  &  $-$0.00343  &  $-$0.00011  &  PE   &  I   &  B\&C (1981a)         \\
43,543.55006   &  $\pm$0.00004  &  27,141.0  &  $-$0.00330  &  $-$0.00001  &  PE   &  I   &  B\&C (1981a)         \\
43,544.5411    &  $\pm$0.0005   &  27,142.5  &  $-$0.00143  &     0.00186  &  PE   &  II  &  B\&C (1981a)         \\
43,577.5120    &  $\pm$0.0004   &  27,192.5  &  $-$0.00274  &     0.00051  &  PE   &  II  &  B\&C (1981a)         \\
43,589.3798    &  $\pm$0.0010   &  27,210.5  &  $-$0.00494  &  $-$0.00170  &  PE   &  II  &  B\&C (1981a)         \\
43,849.53175   &  $\pm$0.00006  &  27,605.0  &  $-$0.00378  &  $-$0.00080  &  PE   &  I   &  B\&C (1981a)         \\
43,876.56881   &  $\pm$0.00006  &  27,646.0  &  $-$0.00394  &  $-$0.00098  &  PE   &  I   &  B\&C (1981a)         \\
43,905.58451   &  $\pm$0.00001  &  27,690.0  &  $-$0.00379  &  $-$0.00086  &  PE   &  I   &  B\&C (1981a)         \\
43,926.3566    &  $\pm$0.0003   &  27,721.5  &  $-$0.00419  &  $-$0.00129  &  PE   &  II  &  B\&C (1981a)         \\
51,650.7741    &  $\pm$0.0001   &  39,435.0  &     0.01215  &     0.00172  &  CCD  &  I   &  Nelson (2001)        \\
52,297.0294    &  \dots         &  40,415.0  &     0.01200  &  $-$0.00001  &  CCD  &  I   &  Nagai (2003)         \\
52,585.8672    &  $\pm$0.0003   &  40,853.0  &     0.01319  &     0.00044  &  CCD  &  I   &  Nelson (2003)        \\
52,628.7304    &  $\pm$0.0002   &  40,918.0  &     0.01251  &  $-$0.00036  &  CCD  &  I   &  Dvorak (2003)        \\
52,912.2932    &  \dots         &  41,348.0  &     0.01424  &     0.00064  &  CCD  &  I   &  Nagai (2004)         \\
52,976.2574    &  \dots         &  41,445.0  &     0.01234  &  $-$0.00142  &  CCD  &  I   &  Nagai (2004)         \\
53,677.9092    &  $\pm$0.0001   &  42,509.0  &     0.01538  &  $-$0.00029  &  CCD  &  I   &  Dvorak (2006)        \\
53,679.2286    &  \dots         &  42,511.0  &     0.01589  &     0.00022  &  CCD  &  I   &  Nagai (2006)         \\
53,683.1850    &  \dots         &  42,517.0  &     0.01562  &  $-$0.00006  &  CCD  &  I   &  Nagai (2006)         \\
53,730.9978    &  \dots         &  42,589.5  &     0.01871  &     0.00290  &  CCD  &  II  &  Nagai (2006)         \\
54,085.4471    &  $\pm$0.0002   &  43,127.0  &     0.01668  &  $-$0.00012  &  PE   &  I   &  H\"ubscher (2007)    \\
54,127.6524    &  $\pm$0.0003   &  43,191.0  &     0.01754  &     0.00062  &  CCD  &  I   &  Ogloza et al. (2008) \\
54,188.3217    &  $\pm$0.0013   &  43,283.0  &     0.01796  &     0.00087  &  CCD  &  I   &  Br\'at et al. (2007) \\
54,461.33036   &  $\pm$0.00036  &  43,697.0  &     0.01667  &  $-$0.00121  &  CCD  &  I   &  This paper           \\
54,497.27047   &  $\pm$0.00035  &  43,751.5  &     0.01707  &  $-$0.00092  &  CCD  &  II  &  This paper           \\
54,512.10740   &  $\pm$0.00003  &  43,774.0  &     0.01650  &  $-$0.00153  &  CCD  &  I   &  This paper           \\
54,523.97755   &  $\pm$0.00007  &  43,792.0  &     0.01665  &  $-$0.00141  &  CCD  &  I   &  This paper
\enddata
\end{deluxetable}

\begin{deluxetable}{lccc}
\tablewidth{0pt}
\tablecaption{Light-curve sets for GW Gem.}
\tablehead{
\colhead{Reference}          & \colhead{Season} & \colhead{Data type} & \colhead{$\sigma$$\rm ^a$} }
\startdata
B\&C                         & 1978-1979        & $B$                 & 0.0083            \\
                             &                  & $V$                 & 0.0075            \\
SOAO                         & 2007-2008        & $B$                 & 0.0124            \\
                             &                  & $V$                 & 0.0100            \\
                             &                  & $R$                 & 0.0113            \\
                             &                  & $I$                 & 0.0088            \\
\enddata
\tablenotetext{a}{In units of total light at phase 0.75.}
\end{deluxetable}

\begin{deluxetable}{ccc}
\tablewidth{0pt}
\tablecaption{GW Gem parameters determined by analyzing all light curves simultaneously$\rm ^a$.}
\tablehead{
\colhead{Parameter}             & \colhead{Primary}        & \colhead{Secondary}
}
\startdata
$T_0$ (HJD)                     & \multicolumn{2}{c}{2425645.5826$\pm$0.0030}              \\
$P$ (d)                         & \multicolumn{2}{c}{0.65944290$\pm$0.00000018}            \\
d$P$/d$t$                       & \multicolumn{2}{c}{(1.130$\pm$0.075)$\times$10$^{-10}$}  \\
$q$                             & \multicolumn{2}{c}{0.460$\pm$0.002}                      \\
$i$ (deg)                       & \multicolumn{2}{c}{81.60$\pm$0.06}                       \\
$T$ (K)                         & 7700                     & 5007$\pm$7                    \\
$\Omega$                        & 3.285$\pm$0.006          & 2.799                         \\
Fill-out (\%)                   & 85.2                     & 100                           \\
$X$, $Y$                        & 0.670, 0.210             & 0.642, 0.165                  \\
$x_{B}$, $y_{B}$                & 0.818, 0.328             & 0.853, $-$0.016               \\
$x_{V}$, $y_{V}$                & 0.712, 0.284             & 0.799, 0.117                  \\
$x_{R}$, $y_{R}$                & 0.592, 0.249             & 0.714, 0.176                  \\
$x_{I}$, $y_{I}$                & 0.483, 0.225             & 0.617, 0.195                  \\
$L/(L_1+L_2)_{B}$               & 0.9482(7)                & 0.0518                        \\
$L/(L_1+L_2)_{V}$               & 0.9072(8)                & 0.0928                        \\
$L/(L_1+L_2)_{B}$               & 0.9482(9)                & 0.0518                        \\
$L/(L_1+L_2)_{V}$               & 0.9072(10)               & 0.0928                        \\
$L/(L_1+L_2)_{R}$               & 0.8658(11)               & 0.1334                        \\
$L/(L_1+L_2)_{I}$               & 0.8255(11)               & 0.1745                        \\
$r$ (pole)                      & 0.3508(7)                & 0.2933(4)                     \\
$r$ (point)                     & 0.3835(11)               & 0.4210(16)                    \\
$r$ (side)                      & 0.3628(8)                & 0.3060(4)                     \\
$r$ (back)                      & 0.3737(10)               & 0.3386(4)                     \\
$r$ (volume)$\rm ^b$            & 0.3628                   & 0.3139                        \\
\enddata
\tablenotetext{a}{Bandpass-specific luminosities are listed in the same order as the entries in Table 3.}
\tablenotetext{b}{Mean volume radius.}
\end{deluxetable}

\begin{deluxetable}{cccccc}
\tablewidth{0pt}
\tablecaption{GW Gem parameters obtained from SOAO light curves.}
\tablehead{
\colhead{Parameter}                 & \multicolumn{2}{c}{Hot-Spot Model}       && \multicolumn{2}{c}{Cool-Spot Model}      \\ [1.0mm] \cline{2-3} \cline{5-6} \\ [-2.0ex]
                                    & \colhead{Primary}  & \colhead{Secondary} && \colhead{Primary}  & \colhead{Secondary}
}
\startdata
$\phi$$\rm ^a$                      & \multicolumn{2}{c}{0.00014(7)}           && \multicolumn{2}{c}{$-$0.00023(7)}        \\
$q$                                 & \multicolumn{2}{c}{0.459(2)}             && \multicolumn{2}{c}{0.458(2)}             \\
$i$ (deg)                           & \multicolumn{2}{c}{81.84(4)}             && \multicolumn{2}{c}{81.98(4)}             \\
$T$ (K)                             & 7700               & 5004(6)             && 7700               & 5004(6)             \\
$\Omega$                            & 3.276(4)           & 2.797               && 3.257(4)           & 2.795               \\
Fill-out (\%)                       & 85.7               & 100                 && 86.1               & 100                 \\
$x_{B}$                             & 0.847(17)          & 0.486(23)           && 0.846(17)          & 0.552(23)           \\
$x_{V}$                             & 0.740(18)          & 0.722(11)           && 0.745(17)          & 0.736(11)           \\
$x_{R}$                             & 0.628(23)          & 0.772(9)            && 0.646(23)          & 0.744(9)            \\
$x_{I}$                             & 0.471(23)          & 0.615(6)            && 0.490(22)          & 0.583(6)            \\
$L/(L_1+L_2)_{B}$                   & 0.940(1)           & 0.060               && 0.942(1)           & 0.058               \\
$L/(L_1+L_2)_{V}$                   & 0.904(1)           & 0.096               && 0.906(1)           & 0.094               \\
$L/(L_1+L_2)_{R}$                   & 0.868(1)           & 0.132               && 0.868(1)           & 0.132               \\
$L/(L_1+L_2)_{I}$                   & 0.827(1)           & 0.173               && 0.827(1)           & 0.173               \\
$r$ (pole)                          & 0.3518(15)         & 0.2932(7)           && 0.3540(16)         & 0.2930(8)           \\
$r$ (point)                         & 0.3850(24)         & 0.4208(31)          && 0.3882(26)         & 0.4205(33)          \\
$r$ (side)                          & 0.3640(18)         & 0.3058(8)           && 0.3665(19)         & 0.3056(8)           \\
$r$ (back)                          & 0.3750(20)         & 0.3384(8)           && 0.3778(22)         & 0.3382(8)           \\
$r$ (volume)                        & 0.3639             & 0.3137              && 0.3664             & 0.3135              \\[1.0mm]
Colatitude (deg)                    & 86.3(7)            & \dots               && \dots              & 80.3(2.4)           \\
Longitude (deg)                     & 123.2(7)           & \dots               && \dots              & 112.7(3.7)          \\
Radius (deg)                        & 10.2(4)            & \dots               && \dots              & 16.2(1.4)           \\
$T$$\rm _{spot}$/$T$$\rm _{local}$  & 1.060(6)           & \dots               && \dots              & 0.754(32)           \\
$\Sigma W(O-C)^2$                   & \multicolumn{2}{c}{0.0122}               && \multicolumn{2}{c}{0.0121}               \\
\enddata
\tablenotetext{a}{Phase shift from the data phased by the quadratic ephemeris of Table 4.}

\end{deluxetable}

\begin{deluxetable}{ccc}
\tablewidth{0pt}
\tablecaption{Astrophysical data for GW Gem.}
\tablehead{
\colhead{Parameter}  & \colhead{Primary} & \colhead{Secondary}}
\startdata
$M$ (M$_\odot$)      &  1.74         &  0.80           \\
$R$ (R$_\odot$)      &  1.75         &  1.50           \\
$\log$ $g$ (cgs)     &  4.19         &  3.99           \\
$L$ (L$_\odot$)      &  9.65         &  1.26           \\
$M_{\rm bol}$ (mag)  &  +2.23        &  +4.44          \\
BC (mag)             &  $+$0.03      &  $-$0.30        \\
$M_{V}$ (mag)        &  +2.20        & +4.74           \\
Distance (pc)        &  \multicolumn{2}{c}{455}        \\
\enddata
\end{deluxetable}

\clearpage
\begin{figure}
 \includegraphics[]{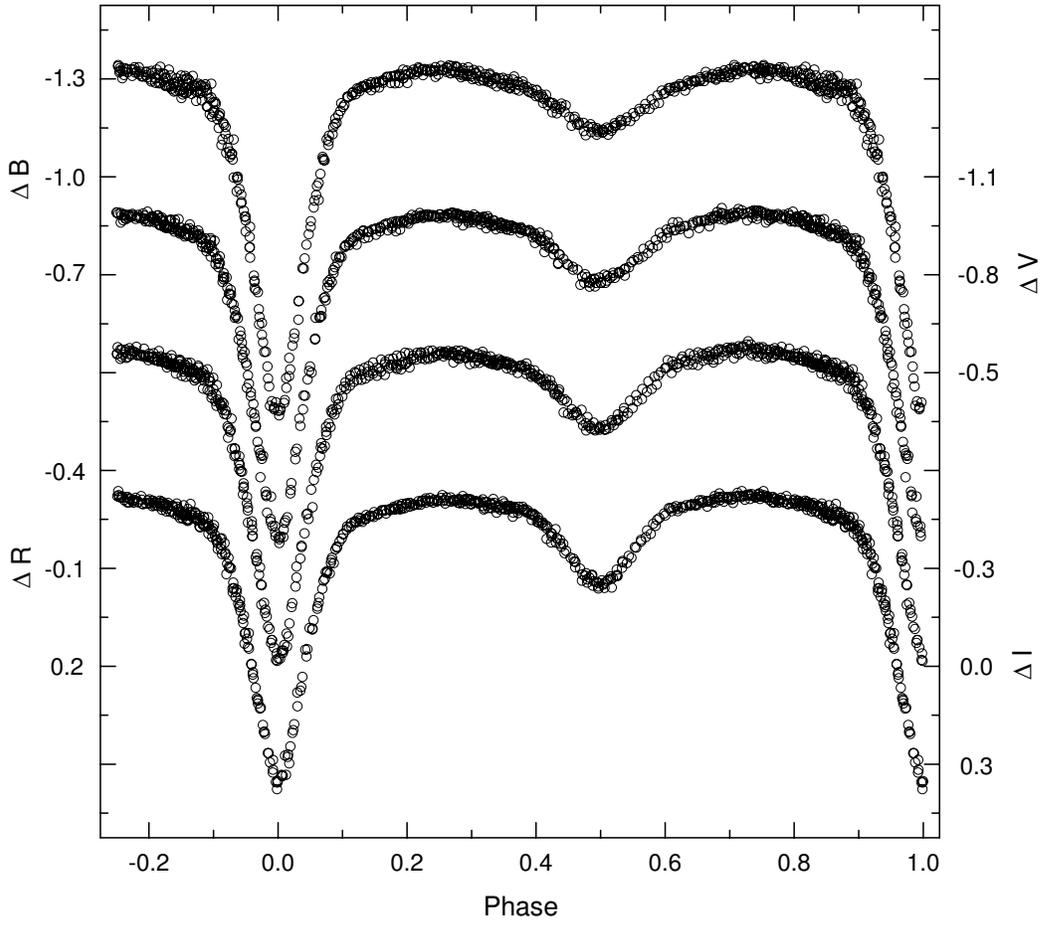}
 \caption{SOAO light curves of GW Gem in the $B$, $V$, $R$, and $I$ bandpasses.}
 \label{f1}
\end{figure}

\begin{figure}
 \includegraphics[]{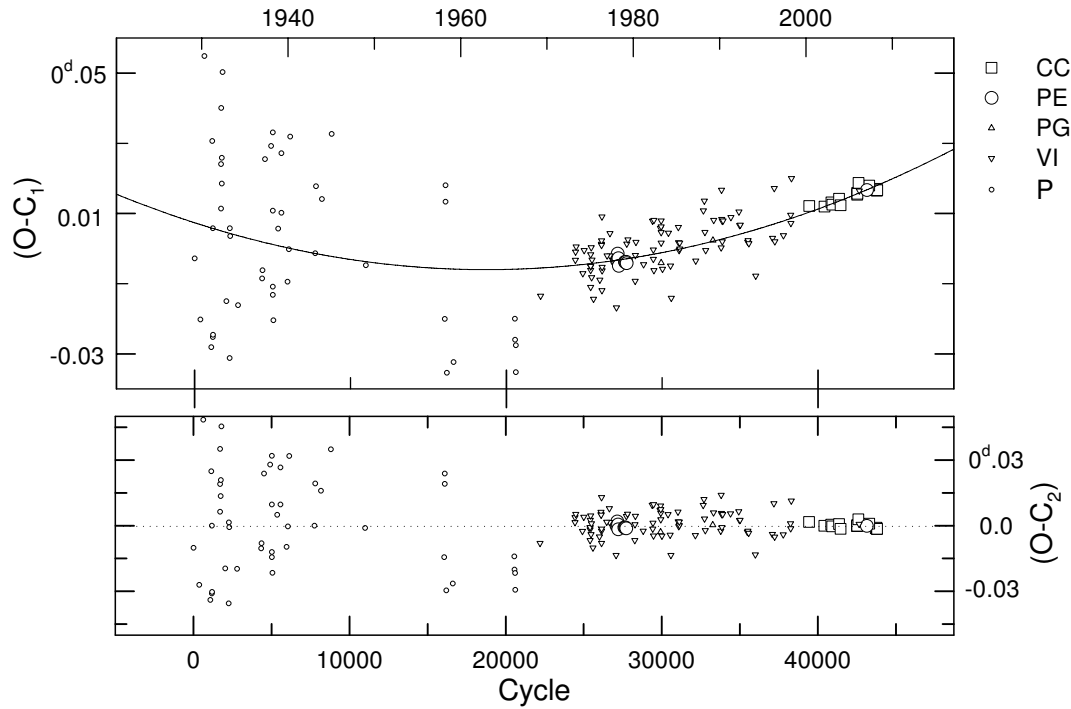}
 \caption{The $O$--$C$ diagram of GW Gem. In the upper panel constructed with the light elements of Kreiner et al.,
 the continuous curve represents the quadratic term of equation (2). The residuals from the quadratic ephemeris
 are plotted in the lower panel. CC, PE, PG, VI, and P represent CCD, photoelectric, photographic, visual,
 and photographic plate minima, respectively. }
 \label{f2}
\end{figure}

\begin{figure}
 \includegraphics[]{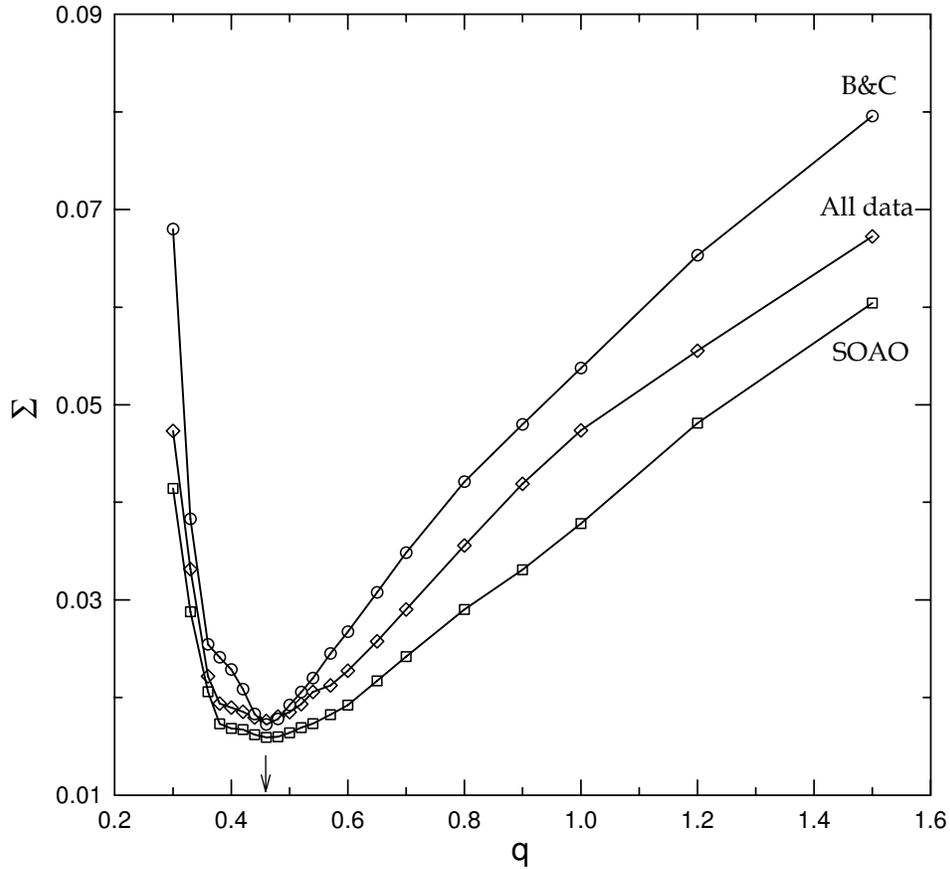}
 \caption{The behavior of $\Sigma$ for GW Gem as a function of assumed mass ratio $q$. The circles, squares, and diamonds
 represent the $q$-search results for B\&C, SOAO, and all data sets, respectively. The arrow indicates
 the minimum value of $\Sigma$ close to $q$=0.46.}
\label{f3}
\end{figure}

\begin{figure}
 \includegraphics[]{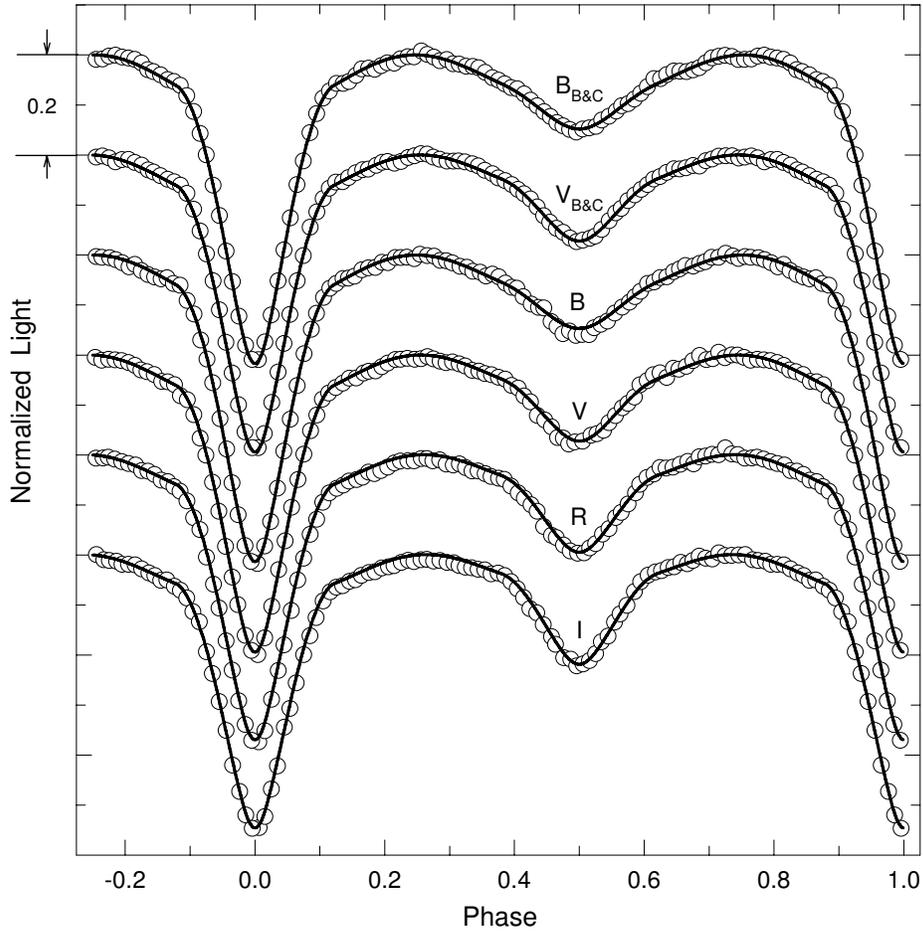}
 \caption{Normalized observations of GW Gem with the theoretical light curves obtained by fitting simultaneously
 all data sets (from top to bottom, the same order as the entries in Table 3). The continuous curves represent
 the solutions obtained with our model parameters listed in Table 4.}
\label{f4}
\end{figure}

\begin{figure}
 \includegraphics[]{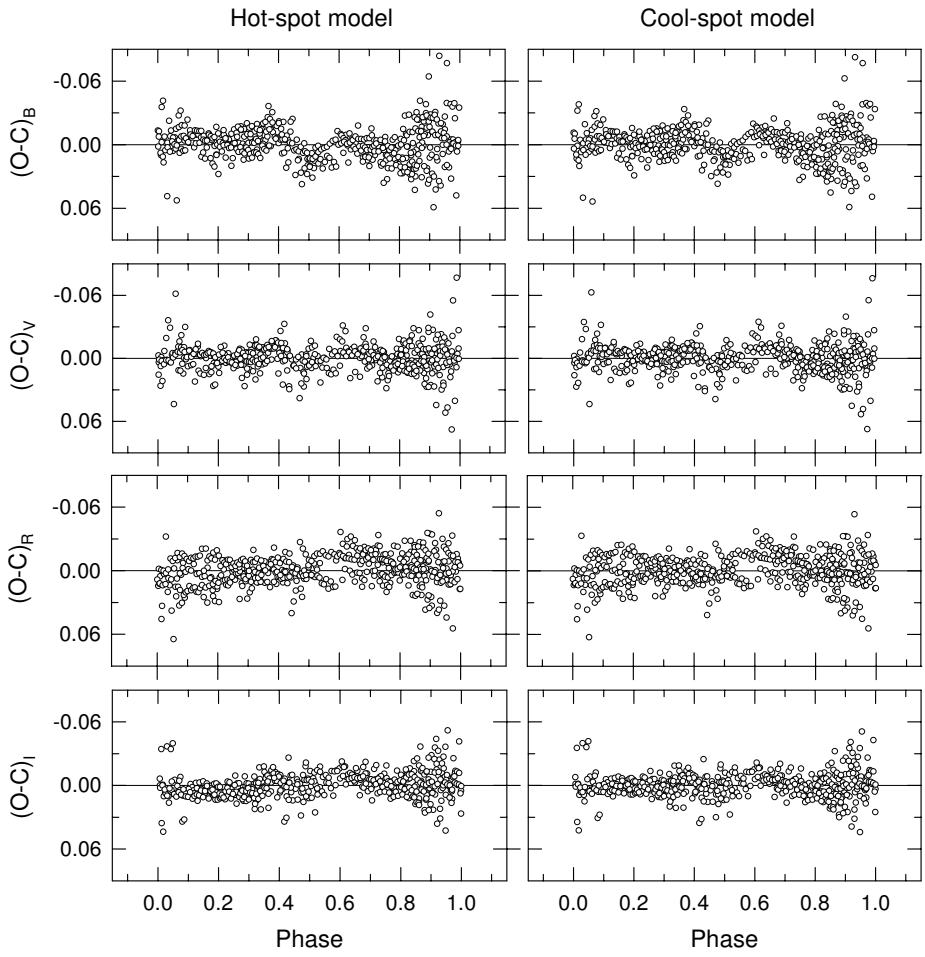}
 \caption{The light residuals of the SOAO curves from the spot models of Table 5.}
\label{f5}
\end{figure}

\end{document}